\documentclass[twocolumn,aps,prl,amsmath,showpacs,amssymb,floatfix]{revtex4}
\usepackage{tabularx}
\usepackage{bm}
\usepackage{subfigure}
\usepackage{euscript}
\usepackage{graphicx}
\usepackage{color}
\usepackage[colorlinks=true,linkcolor=blue]{hyperref}%
\usepackage{amsfonts}
\usepackage{exscale}
\usepackage{amsbsy}

\begin{document}

\title{Long-range Order in One-dimensional Spinless Fermi Gas\\ with Attractive
Dipole-Dipole Interaction}

\author{Zhongbo Yan$^1$}
\author{Liang Chen$^{2}$}
\author{Shaolong Wan$^1$}
\email[]{slwan@ustc.edu.cn}
\affiliation{$^{1}$Institute for Theoretical
Physics and Department of Modern Physics University of Science and
Technology of China, Hefei, 230026, P. R. China\\
$^2$Beijing Computational Science Research Center, Beijing,
100084, P. R. China}
\date{\today}

\begin{abstract}
One-dimensional spinless Fermi gas with attractive dipole-dipole
interaction is investigated. Results obtained show when the
interaction is weak, the excitation spectrum is linear and the
superconducting correlation function decays as power law,
indicating the validity of the Tomonaga-Luttinger (TL) liquid
picture. However, when the interaction reaches a critical value,
the excitation spectrum is nonlinear and the superconducting
correlation function keeps finite for infinity separation,
indicating real long-range order established and the breakdown of
the TL liquid picture. We prove that the existence of long-range
order is not in contradiction with the Hohenberg theorem and show
that this system is related to the Kitaev toy model, therefore, it
has potential applications for the future topological quantum
computation.
\end{abstract}


\maketitle

{\it Introduction.---} The Tomonaga-Luttinger (TL) liquid theory
\cite{S. Tomonaga, J. M. Luttinger, F. D. M. Haldane} can well
describe a lot of one-dimensional physics systems \cite{A. O.
Gogolin, T. Giamarchi}, such as the edge state of a topological
nontrivial system, quantum wires, organic conductors, etc. For
one-dimensional homogeneous spinless fermions, if the fermions are
neutral (cold atomic system), usually the system is described by a
free Fermi gas model, since the two-body collisions are negligible
due to the Pauli principle. If the fermions are charged, like
fully polarized electrons, they feel a long-range Coulumb
repulsive interaction, and the repulsive interaction drives the
system far away from the free Fermi gas to the quasi-Wigner
crystal phase \cite{H. J. Schulz}.

In this paper, we consider one-dimensional spinless fermions with
dipole-dipole interaction in a cold atomic system. The dipolar
interaction has recently arisen a lot of interests in cold atomic
field \cite{L. You, M. A. Baranov1, T. Lahaye} due to its
long-range and anisotropic character. The long-range and
anisotropic character can induce many peculiar effects in both
bosonic and fermionic systems. For bosons, for example, the
dipole-dipole interaction can induce the excitation spectrum of a
Bose-Einstein condensation to exhibit interesting roton structure
\cite{D. H. J. O¡¯Dell, L. Santos, S. Ronen}. For fermions, under
different conditions the dipole-dipole interaction can induce
numerous phases of interest, such as p-wave superfluid \cite{M. A.
Baranov2, G. M. Bruun}, ferronematic phase \cite{B. M. Fregoso},
liquid crystal phase \cite{Chungwei Lin} and other phases.

For neutral spinless fermions in cold atomic system, because the
collisions between fermions are negligible, the dipole-dipole
interaction is the only familiar and available interaction that
can be used to drive the system away from the free Fermi gas.
Before making any progress, we should notice that the
dipole-dipole interaction in one dimension is no longer long-range
but finite range. As a result, if the dipoles are aligned to the
direction perpendicular to the system which indicates that the
interaction is repulsive, no critical behavior like quasi-Wigner
crystal phase emerges and the system is well described by the TL
liquid theory for all interaction strength \cite{Y. Tsukamoto, H.
Inoue}. The purpose of the present paper is to investigate the
effects of the attractive dipole-dipole interaction which is
induced by aligning the dipoles along the one-dimensional system.
The main conclusion we have obtained by bosonization technique is
that the finite-range attractive dipole-dipole interaction, when
its strength reaches a critical value where TL liquid theory
reaches the critical point of validity, establishes a phase
(superfluid) with real long-range order, instead of
quasi-long-range order which is  power-law decay.

{\it Model.---} For one-dimensional homogeneous spinless fermions
with attractive dipole-dipole interaction, the Hamiltonian has the
form (here setting $\hbar=1$)
\begin{eqnarray}
H = \sum_{k;r=\pm} v_{F} (r k - k_{F}) c_{r,k}^{\dag} c_{r,k} +
\frac{1}{2L} \sum_{q} V(q) \rho(q) \rho(-q), \label{1}
\end{eqnarray}
where $r=+$ for right going particles and  $r=-$ for left going
particles. $L$ is the length of the system.
$\rho(q) = \sum_{r,k} c_{r,k+q}^{\dag} c_{r,k}$ is the
density fluctuation operator. We also define $\rho_{r}(q) =
\sum_{k} c_{r,k+q}^{\dag} c_{r,k}$ for convenience. $V(q)$ is the
Fourier form of the dipole-dipole interaction
\begin{eqnarray}
V(x) = - \frac{C_{dd}}{[x^{2} + d^{2}]^{\frac{3}{2}}}, \label{2}
\end{eqnarray}
and takes the form
\begin{eqnarray}
V(q) = - \frac{2C_{dd}}{d^{2}}|q d|K_{1}(q d), \label{3}
\end{eqnarray}
where $K_{1}$ denotes the 1-st order modified Bessel function,
$C_{dd} = \tilde{d}^{2}/2\pi\epsilon_{0}$ for fermions with
electric dipole moment $\tilde{d}$ or $C_{dd} = \mu_{0} \mu^{2}/2
\pi$ for fermions with magnetic dipole moment $\mu$. For the
dipole-dipole interaction, we followed Ref.\cite{H. J. Schulz} and
introduced a cut off $d$ to avoid the divergence at $r=0$. The cut
off $d=\lambda(l/l_{\perp})l_{\perp}$, where
$\lambda(l/l_{\perp})$ is a constant whose value is determined by
the ratio $l/l_{\perp}$, $l$ is the length of the dipole and
$l_{\perp}=\sqrt{\hbar/m\omega_{\perp}}$ is the radial confinement
length. In fact, the simple interaction form (\ref{2}) has
maintained the essential properties of the dipole-dipole
interaction both for long distance and short distance. This can be
easily seen if we deduce the dipole-dipole interaction from the
microscopic structure of the dipole. The moment of the dipole is
defined as $\tilde{d}=ql$, where $q$ is the polarized charge and
$l$ is the length as mentioned above. As charges interact with
each other according to the Coulomb principle, the interaction
between two dipoles is given as
\begin{eqnarray}
V(x) = - \frac{q^{2}}{4\pi\epsilon_{0}}(\frac{1}{x}+\frac{1}{x+2l}-\frac{2}{x+l}), \nonumber
\end{eqnarray}
where $x$ is the distance from the left dipole's tail to the right
dipole's head in the one-dimensional configuration. Such an
interaction form also divergent at $x=0$. To avoid the divergence,
we introduce the transverse confinement $l_{\perp}$ as a cut off
and give
\begin{eqnarray}
V(x) = - \frac{q^{2}}{4\pi\epsilon_{0}}&(&\frac{1}{(x^{2}+
l_{\perp}^{2})^{1/2}}+\frac{1}{((x+2l)^{2}+l_{\perp}^{2})^{1/2}}\nonumber \\
&&-\frac{2}{((x+l)^{2}+l_{\perp}^{2})^{1/2}}). \nonumber
\end{eqnarray}
When $x>>d,l$, by using a Taylor expansion, we obtain
$V(x)=-\frac{C_{dd}}{x^{3}}$. When $x\rightarrow0$ and
$l_{\perp}<<l$, $V(x)=-\frac{C_{dd}}{2l_{\perp}l^{2}}
(1-\frac{l_{\perp}}{l})$. However, when $x\rightarrow0$
and $l_{\perp}>>l$, $V(x)=\frac{C_{dd}}{l_{\perp}^{3}}$, which
is positive. This indicates that the interaction form (\ref{2})
is valid for short distance only when the system is really
very close to be one-dimensional, i.e., $l_{\perp}$ is the
same order of $l$. In this paper, we set $l_{\perp}=l$ for
simplicity, then $V(x=0)=-\frac{C_{dd}}{d^{3}}$, with
$\lambda(l/l_{\perp})\approx60^{\frac{1}{3}}$. Based on
the analysis above, the justification of using (\ref{2})
to substitute the dipole-dipole interaction is evident.

In fact, we more care about the Fourier form of the interaction,
because it determines all quantities we care about. From
(\ref{3}), we can see that both $C_{dd}$ and $d$ can be easily
tuned, therefore the interaction can have a wide range.
Furthermore, to guarantee that the system is effectively
one-dimensional, the density $n$ should satisfy the condition:
$n<1/{l_{\perp}}$.

We rewrite the Hamiltonian (1) into the g-ology description
\begin{eqnarray}
H&=&\frac{\pi
v_{F}}{L}\sum_{p\neq0}\sum_{r=\pm}:\rho_{r}(p)\rho_{r}(-p): \nonumber \\
&&+\frac{2}{L}\sum_{p}g_{2}(p)\rho_{+}(p)\rho_{-}(-p)  \nonumber \\
&&+\frac{1}{L}\sum_{p,r=\pm}g_{4}(p):\rho_{r}(p)\rho_{r}(-p):,\label{4}
\end{eqnarray}
where $g_{2}(p)=g_{4}(p)=V(p)/2$. Here we have neglected some
constant terms which do not affect the physics we consider.

Following standard procedures in the bosonization, the Hamiltonian
(\ref{4}) can be easily diagonalized. The diagonalized Hamiltonian
$H_{D}$ takes the form
\begin{eqnarray}
H_{D}=\frac{\pi}{L}\sum_{p\neq0}\sum_{r=\pm}v_{\rho}(p):\rho_{r}(p)\rho_{r}(-p):.
\label{5}
\end{eqnarray}
Here we also only neglect the constant terms and only keep the
terms related to the physics we care about. In Hamiltonian
(\ref{5}), $v_{\rho}(p)=v_{F}/K_{\rho}(p)=v_{F}\sqrt{1+V(p)/\hbar\pi
v_{F}}$ (based on dimension analysis, here we have reintroduced
$\hbar$ for following use). $K_{\rho}(p)=1/\sqrt{1+V(p)/\hbar\pi v_{F}}$
is the stiffness which determines quantities like different kinds of
correlation functions.

In one dimension, it's known that the $1/r$ long-range Coulomb
repulsive interaction enhances the $4k_{F}$ charge density
correlations for spinful electrons, and drives the system to a
Wigner crystal which shows a critical behavior quite different
from the ordinary TL liquid \cite{H. J. Schulz}. Such a Wigner
crystal phase is due to the fact that the Fourier form of the
$1/r$ long-range repulsive interaction $V_{e}(q)$ is logarithmic
divergent for $q\rightarrow0$. A divergent $V_{e}(q=0)$ means
$K_{\rho}(p=0)=0$ (for small but non-zero momentum, $K_{\rho}(p)$
keeps small but finite). Here a fact that we should notice is that
the interaction is long-range, instead of a short one which can
usually be treated as a delta interaction, the dependence on
momentum of $K_{\rho}$ is important and we can no longer just use
$K_{\rho}(p=0)$ to substitute the whole $K_{\rho}(q)$ when we
calculate quantities like the correlation functions.

For the system with delta-form attractive interaction, it is know
that when the interaction increases, the fluctuation of
superconducting order is enhanced, however, when the interaction
is strong enough to make $K_{\rho}\rightarrow\infty$, instead of
real long-range superconducting order established, the TL liquid
picture breaks down, as the compressibility $\kappa
=2K_{\rho}/n^{2}\pi v_{\rho}$ is divergent and the system is phase
separated \cite{M. Nakamura, D. C. Cabra}. For attractive
dipolar-dipolar interaction, the interaction is finite-range and
the momentum dependence of the Fourier form needs to be
considered. As $K_{\rho}(p=0)^{-1}$ decreases with increasing
interaction, it will touch zero when the interaction reaches a
critical value. However, as the interaction is finite-range, like
the long-range Coulomb interaction, now we can no longer just use
$K_{\rho}(p=0)^{-1}$ to substitute the whole $K_{\rho}(q)^{-1}$
when we calculate the superconducting correlation function.
Because of the momentum dependence, the relation $\kappa
=2K_{\rho}/n^{2}\pi v_{\rho}$ is no longer justified, in other
words, $K_{\rho}(p=0)^{-1}=0$ does not indicate the
compressibility is divergent and the system will be phase
separated.

In statistical mechanics, the compressibility is defined as
$\kappa=n^{-2}(\partial n/\partial\mu)_{T}$. The divergent of
compressibility means that changing the number of the particles,
the chemical potential does not change, $i.e.$,
$\Delta\mu=E(N+1)+E(N-1)-2E(N)=0$. For a TL liquid with system
length $L$, $\Delta\mu$ should be approximately proportional to
$v_{\rho,\rm\scriptscriptstyle N+1}(q_{\rm\scriptscriptstyle
N+1})q_{\rm\scriptscriptstyle N+1}- v_{\rho,\rm\scriptscriptstyle
N}(q_{\rm\scriptscriptstyle N})q_{\rm\scriptscriptstyle N}$, where
$q_{\rm\scriptscriptstyle N+1}=q_{\rm\scriptscriptstyle
N}+\frac{\pi}{L}$. This agrees with the fact that, more generally,
a vanishing $v_{\rho}$, instead of an infinity $K_{\rho}$,
corresponds to the instability of phase separation \cite{D. C.
Cabra}. However, for attractive dipole-dipole interaction, when
the interaction reaches the critical value, the velocity only
vanishes at $q=0$ (where the bosonic operator is also not
defined), and furthermore, here $v_{\rho,\rm\scriptscriptstyle N}$
is anti-proportional to $N$ (see following). This indicates
$\Delta\mu$ may be very small, but not zero. Therefore, when the
interaction reaches the critical value, the system should not be
phase-separated.

In bosonization language, the single particle fermionic operator
is given as
\begin{eqnarray}
\psi_{r}(x)=\lim_{\alpha\rightarrow 0}
\frac{U_{r}}{\sqrt{2\pi\alpha}}e^{irk_{F}x}e^{-i[r\phi(x)-\theta(x)]},
\label{6}
\end{eqnarray}
where $U_{r}$ is known as Klein factor. For repulsive interaction,
$K_{\rho}(p)<1$, the system is inclined to order in charge density
wave and the charge correlation function $<\rho(x)\rho(0)>$ which
is dominated by the quantity given as
\begin{eqnarray}
&&\frac{2}{(2\pi\alpha)^{2}}\cos(2k_{F}x)e^{-2<[\phi_{\rho}(x)-\phi_{\rho}(0)]^{2}>} \nonumber \\
&&=\frac{2}{(2\pi\alpha)^{2}}\cos(2k_{F}x)e^{-2\int_{0}^{\infty}\frac{dp}{p}K_{\rho}(p)(1-\cos px)} \label{7}
\end{eqnarray}
is mainly considered. From (\ref{7}), it is easy to see if we
directly substitute $K_{\rho}(p)$ as $K_{\rho}(p=0)$ which takes
value zero for the long-range repulsive Coulomb interaction, the
correlation function will exhibit perfect crystalline order, which
is a wrong conclusion as shown in Ref.\cite{H. J. Schulz}. This
gives us a concrete example that we can not directly use
$K_{\rho}(p=0)$ to substitute $K_{\rho}(p)$ when the interaction
is strong enough to induce a vanishing $K_{\rho}(p=0)$, or else
wrong conclusion may be obtained.

For attractive interaction, $K_{\rho}(p)>1$,
the system is inclined to order in superconductivity and
what we care about is the superconducting correlation
function which is mainly determined by the quantity given as
\begin{eqnarray}
&&\frac{1}{(\pi\alpha)^{2}}e^{-2<[\theta_{\rho}(x)-\theta_{\rho}(0)]^{2}>} \nonumber \\
&&=\frac{1}{(\pi\alpha)^{2}}e^{-2\int_{0}^{\infty}\frac{dp}{p}
K_{\rho}^{-1}(p)(1-\cos px)}. \label{8}
\end{eqnarray}
The similarity of the form between
$<[\phi_{\rho}(x)-\phi_{\rho}(0)]^{2}>$ and
$<[\theta_{\rho}(x)-\theta_{\rho}(0)]^{2}>$ is just the basis for
our expectation that like the long-range repulsive Coulomb
interaction diving the charge correlation function to exhibit a
quasi-Wigner crystal phase \cite{H. J. Schulz}, the finite-range
attractive dipole-dipole interaction may also drive the
superconducting correlation functions to exhibit some critical
behavior when $K_{\rho}^{-1}(p=0)=0$ (as mentioned above, for
non-zero momentum, $K_{\rho}^{-1}(p)$ needs to be small but finite
to against the phase-separation instability). From (\ref{8}), it
is also easy to see we can not directly substitute
$K_{\rho}^{-1}(p)$ as $K_{\rho}^{-1}(p=0)$ when
$K_{\rho}^{-1}(p=0)$ takes the critical value zero, or else the
superconducting correlation function does not depend on distance,
which is obviously unphysical. For delta-form attractive
interaction,  as $K_{\rho}^{-1}(p)=K_{\rho}^{-1}(p=0)$, when
$K_{\rho}\rightarrow\infty$, the result of the superconducting
correlation function is unphysical, this also indicates that the
TL liquid picture is broken down, and the system is
phase-separated.

When $K_{\rho}^{-1}(p=0)=0$, we find
$V(p)/\hbar\pi v_{F}=-|pd|K_{1}(pd)$ or $2C_{dd}/\hbar\pi v_{F}d^{2}=1$. If
we define a dimensionless parameter $\gamma=2C_{dd}/\hbar\pi
v_{F}d^{2}$, we find, for $\gamma<1$, the dispersion takes the
form $\omega_{\rho}(p)\sim p$ in the long-wavelength limit, which
indicates a TL liquid and for $\gamma=1$,
$\omega_{\rho}(p)\sim|p^{4}\log p|^{\frac{1}{2}}$, the dispersion
is still gapless but no longer linear, indicating that different
physics may emerge. By a direct calculating, we find, for
$\gamma<1$
\begin{eqnarray}
&&\frac{1}{(\pi\alpha)^{2}}e^{-2<[\theta_{\rho}(x)-\theta_{\rho}(0)]^{2}>} \nonumber \\
&&=\frac{1}{(\pi\alpha)^{2}}e^{-2\int_{0}^{1/\alpha}\frac{dp}{p}K_{\rho}^{-1}(p)(1-\cos px)} \nonumber \\
&&\simeq\frac{1}{(\pi\alpha)^{2}}e^{-2\int_{0}^{1/d}\frac{dp}{p}K_{\rho}^{-1}(p=0)(1-\cos px)-2\ln(d/\alpha)} \nonumber \\
&&=\frac{1}{(\pi d)^{2}}(\frac{d}{x})^{2\sqrt{1-\gamma}}.\label{9}
\end{eqnarray}
Here we substitute $K_{\rho}^{-1}(p)$ as $K_{\rho}^{-1}(p=0)$,
since when $K_{\rho}^{-1}(p=0)\neq0$, such a substitution does not
affect the quality behavior (power-law decay) of the
superconducting correlation function. For $\gamma=1$, after taking
the same procedures, we obtain
\begin{eqnarray}
&&\frac{1}{(\pi\alpha)^{2}}e^{-2<[\theta_{\rho}(x)-\theta_{\rho}(0)]^{2}>} \nonumber \\
&&=\frac{1}{(\pi d)^{2}}e^{-2\int_{0}^{1/d}\frac{dp}{p}K_{\rho}^{-1}(p)(1-\cos px)} \nonumber \\
&&>\frac{1}{(\pi d)^{2}}e^{-4\int_{0}^{\frac{1}{d}}\frac{dp}{p}\sqrt{1-|pd|K_{1}(pd)}} \nonumber \\
&&=\frac{1}{(\pi d)^{2}}e^{-4\int_{0}^{1}\frac{dy}{y}\sqrt{1-yK_{1}(y)}} \nonumber \\
&&=\frac{1}{(\pi d)^{2}}e^{-3.56}, \label{10}
\end{eqnarray}
here we have introduced a cutoff $1/d$ for the momentum to avoid
nonphysical divergence. The results above indicates when $\gamma <
1$, the superconducting correlation function exhibits a familiar
power law decay and vanishes for $x\rightarrow\infty$, agreeing
with our previous arguments based on the dispersion that the
system falls into the TL liquid for $\gamma < 1$. While $\gamma =
1$, the superconducting correlation function keeps finite for $x
\rightarrow \infty$, indicating a real long-range order
established and simultaneously the TL liquid
picture reaches the critical point of validity.
This is our main result in this paper.

The existence of long-range superconducting order here does not
contradict with the Hohenberg theorem which strictly roles out
Bose-Einstein condensation and superconductivity at finite
temperature in one and two dimension \cite{P. C. Hohenberg}. We
can prove this by following the procedures used in Ref.\cite{P. C.
Hohenberg}. Before giving the details of proof, we should notice
the fact that here the Cooper pair is composed of two spinless
fermions, the anomalous average $<\psi(x)\psi(x^{'})>$ is the
quantity to characterize the long-range order. After noticing this
difference, we follow the procedures applied in Ref.\cite{P. C.
Hohenberg}. First, we similarly introduce the order parameter
\begin{eqnarray}
\Delta(x)=\int dx^{'}s(x-x^{'})<\psi(x)\psi(x^{'})>, \nonumber
\end{eqnarray}
where $s(x)$ is a smearing function. By Fourier transformation, we
have
\begin{eqnarray}
\Delta(k)&=& \Omega^{-1} \sum_{q} S(q) <c_{k-q}c_{q}>. \label{11}
\end{eqnarray}
We assume the smearing function has inversion symmetry, therefore,
$S(q)=S(-q)$. As a result, $\Delta(0)=0$.

We apply the Bogoliubov inequality to the operators
\begin{eqnarray}
A_{k}&=&i(\partial/\partial t)\rho_{-k}(t); \nonumber \\
B_{k}&=&\sum_{q}S(q)c_{k-q}c_{q}. \label{12}
\end{eqnarray}
The fermion commutation rules yield for $k\neq 0$
\begin{eqnarray}
\frac{1}{\Omega}<[B_{k},\rho_{-k}]>&=&\frac{1}{\Omega}\sum_{q}[S(q)+S(k-q)]<c_{q}c_{-q}>\nonumber \\
&=&\Delta(0)+\eta(k)=\eta(k), \label{13}
\end{eqnarray}
it's easy to see that $\eta(k)$ is antisymmetric, $i.e.$,
$\eta(k)=-\eta(-k)$, which implies $\eta(k)\propto k^{2n+1}$
(n=0,1,...) for $k\rightarrow0$. Now the Bogoliubov inequality is
given as
\begin{eqnarray}
C_{B,B^{\dag}}(k)\geq2T\frac{|\eta(k)|^{2}}{(n/m)k^{2}}.\label{14}
\end{eqnarray}
The concrete form of $C_{B,B^{\dag}}(k)$ is given as
\begin{eqnarray}
C_{B,B^{\dag}}(k)&=&\Omega^{-1}<\{\sum_{p}S(p)c_{k-p}c_{p},\sum_{q}S(q)c_{q}^{\dag}c_{k-q}^{\dag}\}> \nonumber \\
&=&2\Omega^{-1}\sum_{p,q}S(p)S(q)<c_{q}^{\dag}c_{k-q}^{\dag}c_{k-p}c_{p}>\nonumber \\
&&-\Omega^{-1}\sum_{q}[S(q)-S(k-q)]^{2}<c_{q}^{\dag}c_{q}> \nonumber \\
&&+\Omega^{-1}\sum_{q}S(q)(S(q)-S(k-q))\nonumber \\
&=&F(k)+R(k).\label{15}
\end{eqnarray}
where $F(k)$ denotes the first term in the second line, and $R(k)$
denotes the rest of terms and is regular at small $k$. The Fourier
transform of $F(k)$ is given as
\begin{eqnarray}
f(x_{2}-x_{1})\equiv&&\int\int dx^{'}dx^{''}s(x_{1}-x^{'})s(x_{2}-x^{''}) \nonumber \\
&&\times<\psi^{\dag}(x^{'})\psi^{\dag}(x_{1})\psi(x_{2})\psi(x^{''})>. \label{16}
\end{eqnarray}
By using the conclusion of Ref.\cite{P. C. Hohenberg}, we obtain
\begin{eqnarray}
f(0)\leq 2[\int
dx^{'}s(x_{1}-x^{'})\{<\rho(x_{1})\rho(x^{'})>\}]^{2}. \label{17}
\end{eqnarray}
As pointed by Hohenberg, $f(0)$ is finite since  the density
correlation function may not have any singularities which are not
integrable. Combining Eqs.(\ref{14}), (\ref{15}) and (\ref{17}),
we finally have
\begin{eqnarray}
F(k)\geq2T\frac{|\eta(k)|^{2}}{(n/m)k^{2}}-R(k),\label{18} \\
\Omega^{-1}\sum_{k\neq0}F(k)<f(0)<\infty. \label{19}
\end{eqnarray}
In Ref.\cite{P. C. Hohenberg}, since $\Delta+\eta(k)\rightarrow2\Delta$ and
$\Delta$ is a non-zero constant, Eqs.(\ref{18}) and (\ref{19}) are in contraction in one
and two dimension for $T \neq 0$ and for infinite volume. However,
for spinless fermions, $\eta(k) \sim k^{2n+1}$ (n=0,1,...) for $k \rightarrow
0$, as a result, the infrared divergence is removed and
$F(k)$ is finite. The contraction appearing in Ref.\cite{P. C. Hohenberg}
disappears and Eq.(\ref{18}) and Eq.(\ref{19}) are now self-consistent.
$<c_{q}c_{-q}>$  also no longer needs to vanish for any $q$.
From this, we have proved that the existence
of long-range superconducting order for spinless fermions with
attractive dipolar interaction does not have any contradiction
with the Hohenberg theorem.

For spinful fermions, however, there is no such critical behavior.
This is due to the fact that spin and charge degrees of spinful
fermions will separate in one dimension, which is illustrated by
$K_{\rho}(p)\neq K_{\sigma}(p)$. Furthermore, it is not hard to
find when $K_{\rho}(p)\rightarrow\infty$, $K_{\sigma}(p)$ and
$K_{\sigma}^{-1}(p)$ keep finite. As a result, the superconducting
correlation functions proportional to
$e^{-\int_{0}^{\infty}\frac{dp}{p}(K_{\rho}^{-1}(p)+K_{\sigma}(p))(1-\cos
px)}$ or
$e^{-\int_{0}^{\infty}\frac{dp}{p}(K_{\rho}^{-1}(p)+K_{\sigma}^{-1}(p))(1-\cos
px)}$ always give a power law and therefore always vanishes for $x
\rightarrow \infty$. No real long-range order exists at least when
the attractive interaction is not strong enough to go beyond the
Luttinger liquid picture.

{\it Related to the Kitaev toy model.---}
As the dipole-dipole interaction is not truly long-range in one
dimension, the collisions between spinless fermions is dominated
by $p$-wave collisions at low temperature. Therefore, based on the
arguments in the previous section, when real long-range order is established for
sufficiently strong attractive interaction, the system is a real
$p$-wave superfluid and it is a realization of the Kitaev toy model \cite{A. Yu.
Kitaev}. This suggests such a system has potential applications  for the future
topological quantum computation.

{\it Experiments.---}To observe the critical behavior, we already know
that the interaction should reach the critical value, i.e.,
$\gamma_{c}=2C_{dd}/\hbar\pi v_{F}d^{2}=(2\tilde{d}^{2}m)/
(2\pi^{3}\epsilon_{0}\hbar^{2}\lambda^{2}l_{\perp}^{2}n)=1$.
Assuming $n=0.25/l_{\perp}$, then for polar molecules composed
of alkali atoms \cite{K.-K. Ni}, $m\sim100u$ ($1u=1.66\times10^{-27}kg$),
$l_{\perp}=l\sim10\AA$ (corresponds to $\omega_{\perp}\sim600MHz$),
we obtain the critical $\tilde{d}_{c}=(\pi^{3}\epsilon_{0}
\hbar^{2}\lambda^{2}l_{\perp}^{2}n/m)^{1/2}=0.075$Debye,
which is within the current experimental ability.
From the expression of $d_{c}$, it's easy to see that a lower density $n$
corresponds to a lower $d_{c}$, and a lower $d_{c}$ corresponds to a weaker
interaction, therefore, the three-body loss is not worthy of worry.

In future, once the
polar molecules can be cooled to quantum degeneracy, the system proposed
in this paper is realizable under current experimental conditions and therefore
its applications for topological quantum computation is worth expecting.

{\it Conclusion.---}Using bosonization technique, both the excitation
spectrum and the superconducting correlation function of
one-dimensional spinless fermi gas with attractive dipolar
interaction are obtained. The superconducting correlation function
exhibits a critical behavior that when the interaction reaches the
critical value, $i.e.$ $\gamma=1$, it keeps finite for infinity
separation, indicating real long-range order established. The
existence of this long-range order has no contradiction with the
Hohenberg theorem and makes the system to be a realization of
Kitaev toy model.

{\it Acknowledgments.---}This work is supported by NSFC. Grant No.11275180.

\end{document}